\documentclass[conference]{IEEEtran}
\IEEEoverridecommandlockouts
\pdfoutput=1
\usepackage{citesort}
\usepackage{amsmath,amssymb,amsfonts}
\usepackage{breqn}
\usepackage{algorithmic}
\usepackage{graphicx}
\usepackage{pstool}
\usepackage{psfrag}
\newcommand{\R}{\mathbb{R}}
\usepackage{textcomp}
\usepackage{xcolor}
\def\BibTeX{{\rm B\kern-.05em{\sc i\kern-.025em b}\kern-.08em
    T\kern-.1667em\lower.7ex\hbox{E}\kern-.125emX}}
\begin{document}
\thispagestyle{plain}
\pagestyle{plain}
\title{A new measure of modularity density for community detection\\
\thanks{The authors are with the Data Science group at CKM Analytix, New York City, NY 10036 USA (email: smula@ckmanalytix.com; gveltri@ckmanalytix.com)}
}

\author{\IEEEauthorblockN{\textbf{Swathi M. Mula~~\&~~Gerardo Veltri}}
\textit{CKM Analytix, USA}\\
}

\maketitle

\begin{abstract}
Using an intuitive concept of what constitutes a meaningful community, a novel metric is formulated for detecting non-overlapping communities in undirected, weighted heterogeneous networks.  This metric, modularity density, is shown to be superior to the versions of modularity density in present literature. Compared to the previous versions of modularity density, maximization of our metric is proven to be free from bias and better detect weakly-separated communities particularly in heterogeneous networks.  In addition to these characteristics, the computational running time of our modularity density is found to be on par or faster than that of the previous variants.  Our findings further reveal that community detection by maximization of our metric is mathematically related to partitioning a network by minimization of the normalized cut criterion.
\end{abstract}


\section{Introduction}\label{sec:intro}
Community detection has numerous applications to a variety of practical problems associated with biological, social and internet systems, in which particular systems are represented using networks (graphs) with nodes and edges.  Due to the lack of a standard mathematical definition for what a community is, finding a meaningful community structure in a network is often challenging.  In an intuitive sense, a community can be defined as a group of nodes with dense internal connections (cohesion) and weak external relationships (separation) with nodes of other groups.  One of the popular community detection methods developed on the above concept of cluster cohesion and separation is Shi \& Mallik (2000)'s~\cite{s00} normalized cut approach.  This method divides a network into a set of clusters by minimization of a metric called the normalized cut criterion.  This metric is an unbiased measure of the dissociation between groups as well as the association within groups; however, because the normalized cut criterion requires knowledge of the number of communities in a network prior to its maximization, it is not applicable in cases where one might want to data mine the number of clusters as well as the clusters themselves.  Another popular metric that qualifies the strength of a community structure is modularity, which was originally introduced for undirected, unweighted graphs by {Newman \& Girvan (2004)~\cite{ng04}} and extended to weighted graphs by Newman (2004)~\cite{n04b}.  This metric is grounded in the concept that a random network has no communities within.  This modularity quantifies the deviation of the community structure of a network from that of a random network (null model), which has the same degree sequence as that of the original network.  Since the inception of modularity, numerous community detection algorithms, such as greedy algorithm~\cite{n04a}, simulated annealing~\cite{g04}, spectral optimization~\cite{n06}, genetic algorithm~\cite{t07}, fine-tuned algorithm~\cite{c14}, \emph{etc.}, are introduced based on the maximization of modularity.  Unlike the normalized cut approach~\cite{s00}, modularity-based community detection does not require information on the number of communities prior to the clustering of the network. 

Despite the popularity of modularity-based community detection, optimization of modularity is shown to have some limitations~\cite{f07,g10}.  Investigations by Fortunato \& Barth\'{e}lemy (2007)~\cite{f07} reveal that the modularity-based maximization suffers from the resolution limit problem, \emph{i.e.} the inability to identify clusters smaller than a certain network-specific topological scale.  Besides the resolution limit problem, modularity also suffers from extreme degeneracies and lacks a clear global maximum~\cite{g10}.  To solve the resolution limit problem, Reichardt \& Bornholdt (2006)~\cite{r06} and Arenas \emph{et al.} (2008)~\cite{a08} have proposed different multiresolution variants of modularity.  These versions allow community detection at multiple topological scales. However, as shown by Lancichinetti \& Fortunato (2011)~\cite{l11}, optimization of multiresolution modularity suffers from the two opposite problems of bias: \emph{the tendency to favor smaller clusters over larger ones at high topological resolution and the tendency to favor larger clusters over smaller ones (resolution limit problem) at low resolution}.  This means optimization of multiresolution modularity may fail to detect all the communities in a network especially when the network has a wide range of community sizes, which is most often the case in real-world networks.  Such networks with a wide distribution of community sizes are called heterogenous networks~\cite{d06}. Mathematically, these networks are characterized by power-law distributions of community sizes~\cite{lfr08}.  To enable community detection in heterogeneous networks by optimization of a desired metric, the metric should be free from bias.  

In order to provide a quantitative function superior to modularity, Li \emph{et al.} (2008)~\cite{l08} proposed a new metric called modularity density, which is based on the average modular degree and is equivalent to the objective function of kernel $k$ means.  Optimization of this modularity density does not suffer from the above mentioned problems of bias, \emph{i.e.} the tendency of favoring larger clusters or smaller clusters.  Investigations by Chen \emph{et al.} (2013)~\cite{c13} have also introduced another version of modularity density, which incorporates additional components, such as split penalty and community density, into the mathematical expression of modularity.  While Chen \emph{et al.} (2013)'s~\cite{c13} modularity density performs better than the modularity metric in detecting communities of heterogeneous networks, optimization of this modularity density is shown to still suffer from the resolution limit problem.  To fix this, Chen \emph{et al.} (2018)~\cite{c18} has introduced a new variant of Chen \emph{et al.} (2013)'s~\cite{c13} modularity density; however, the newer version does not completely eliminate the resolution limit problem.  

The objective of the current article is to introduce a new quantitative function that is superior to all the above existing versions of modularity density and enable better community detection in heterogenous networks.  We define our new quantitative function, like Li \emph{et al.} (2008)'s~\cite{l08} and Chen \emph{et al.} (2013)'s~\cite{c13} metrics, as modularity density and formulate this metric using the concept of cohesion and separation.  We show that the optimization of our metric is not only free from bias but also better detects weakly separated communities in heterogeneous networks compared to that of the previous versions of modularity density.

The outline of the remainder of this article is as follows. Using vector and tensor algebra, the new quantitative function is formulated in section~\ref{sec:definition}. To show that our metric is free from bias, mathematical proofs are presented in section~\ref{subsec:bias}.  Comparisons between our metric and the previous versions of modularity density are discussed in section~\ref{sec:comparse}. Finally, mathematical relations between optimization of our metric and the normalized cut approach~\cite{s00} are provided in section~\ref{sec:normcut}, with a summary of our results in section~\ref{sec:conclusion}.

\section{Modularity Density}\label{sec:definition}

\subsection{New quantitative function of modularity density}
Consider an undirected graph $G(V, E)$ comprising a set of vertices $V$ and a set of edges $E$. Let a second-order tensor ${\bold{T} = [T_{ij}]\in {\R}^{|V| \times |V|}}$ represent the adjacency matrix of the network such that ${T_{ij}}$ indicates the weight of the edge  (${i,~j}$). If $C$ is a set of all communities in the network, then each cluster ${c \in C}$ can be represented by an indicator vector ${\vec{v}_c = [v{_{c_i}}] \in {\R}^{|V|} : v{_{c_i}}= 1}$ if ${i \in c}$, else $0$. Using ${\vec{v}_c}$, a unit vector representation of the cluster $c$ is given by 
\begin{IEEEeqnarray}{lll} 
 \hat{n}_c = \frac{\vec{v}_c}{|\vec{v}_c|}, \nonumber
 \end{IEEEeqnarray}  
where $|\vec{v}_c|$ is the Euclidean norm of $\vec{v}_c$. If $n_c$ is the number of nodes in cluster $c$, then $|\vec{v}_c|^2 = n_c$. Therefore,
\begin{IEEEeqnarray}{lll} 
\hat{n}_c = [n_{c_j}] = \frac{\vec{v}_c}{\sqrt{{n}_c}};~n_{c_j} = \begin{cases}
{\frac{1}{\sqrt{n_c}}} \hspace{1.0 em} $~if~$ j \in c\\ 
0 \hspace{2.6 em}  $else$. \\ 
\end{cases}\label{eq:unitvector}
 \end{IEEEeqnarray}   
Interestingly, the dot product of $\hat{n}_c$ and the tensor $\bold{T}$ introduces a new vector $\vec{d}_c \in {\R}^{|V|}$:
\begin{IEEEeqnarray}{lll} 
 \vec{d}_c = [d_{c_j}] = \hat{n}_c \cdot \bold{T}, \label{eq:degree1}\\
 \hspace{2.8em} d_{c_j} = \sum_{i}n_{c_i}T_{ij} = \frac{\sum_{i\in c}T_{ij}}{\sqrt{n_c}}\label{eq:degree2}.
 \end{IEEEeqnarray}  
Note that in the above equation ${\sum_{i\in c}T_{ij}}$ is the sum of the weights of all edges ($i,~j$) connecting nodes $i \in c$ with node $j \in V$.  As presented in equation~(\ref{eq:degree2}), normalizing ${\sum_{i\in c}T_{ij}}$ by ${\sqrt{n_c}}$ generates $d_{c_j}$.  In simple terms, $d_{c_j}$ indicates how the node $j \in V$ is associated with the nodes of cluster $c$.  Following this interpretation of $d_{c_j}$, we define $\vec{d}_c$ in equation~(\ref{eq:degree1}) as a normalized degree vector of the graph  $G$ with respect to cluster~$c$. 
\begin{figure}[htbp]
\psfrag{m}{$it~works$}
\hspace{0.0mm}\centerline{\includegraphics[width=1.3in]{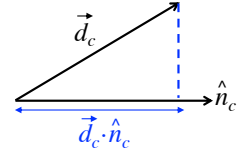}}
\vspace{-7.5mm}
\caption{Projecting the normalized degree vector $\vec{d}_c$ on the unit vector $\hat{n}_c$ of cluster $c$.}
\label{fig:dotproduct}
\end{figure}

To obtain a measure of internal associations (cohesion) within cluster $c$, we take the concept of $\vec{d}_c$ further by projecting this vector on $\hat{n}_c$, which is a unit vector representing cluster $c$ as in (\ref{eq:unitvector}). This projection, as illustrated in figure~\ref{fig:dotproduct}, is determined by the dot product:
\begin{IEEEeqnarray}{lll} 
 \vec{d}_c \cdot \hat{n}_c = \sum_{j}d_{c_j}n_{c_j} = \sum_{j \in c}\frac{d_{c_j}}{\sqrt{n_c}}.
 \label{eq:dotproduct1}
 \end{IEEEeqnarray}  
Using equation~(\ref{eq:degree2}) in~(\ref{eq:dotproduct1}), 
\begin{IEEEeqnarray}{lll} 
 \vec{d}_c \cdot \hat{n}_c =  \frac{1}{\sqrt{n_c}}\sum_{j \in c} \sum_{i \in c} \frac{T_{ij}}{{\sqrt{n_c}}} =  \frac{1}{{n_c}}{\sum_{i, j~\in c} T_{ij}}, 
 \label{eq:internaldensity}
 \end{IEEEeqnarray} 
which indicates that ${\vec{d}_c \cdot \hat{n}_c}$ is equal to the mean internal degree of the cluster $c$.  Therefore, in intuitive terms, ${\vec{d}_c \cdot \hat{n}_c}$ is a measure of the internal associations within cluster $c$.  Likewise, to obtain external associations (separation) between cluster $c$ and cluster $c^{\prime} \in C-c$, we project $\vec{d}_c$ on the corresponding unit vector of ${c^{\prime}}$.  If $\hat{n}_{c^{\prime}}$ and ${n}_{c^{\prime}}$ indicate the unit vector and the number of nodes, respectively, of cluster ${c^{\prime}}$, then we obtain the above projection as:  
\begin{IEEEeqnarray}{lll} 
 \vec{d}_c \cdot \hat{n}_{c^{\prime}} = \sum_{j}d_{c_j}n_{{c_j^{\prime}}} = \sum_{j \in {c^{\prime}}}\frac{d_{c_j}}{\sqrt{n_{c^{\prime}}}} \nonumber \\
 \hspace{3.2em}= \frac{1}{\sqrt{n_{c^{\prime}}}}\sum_{j \in c^{\prime}} \sum_{i \in c}\frac{T_{ij}}{\sqrt{n_c}} = \frac{1}{\sqrt{n_c n_{c^{\prime}}}}{\sum_{i \in c,~ j\in c^{\prime}} T_{ij}}. 
\label{eq:penalty}
\end{IEEEeqnarray} 
In the above equation, $\sum_{i \in c,~ j\in c^{\prime}} T_{ij}$ is the sum of the weights of all edges between clusters $c$ and $c^{\prime}$.  Mathematically, ${\vec{d}_c \cdot \hat{n}_{c^{\prime}}}$ in~(\ref{eq:penalty}) is a normalized measure of the external degree of cluster $c$ with respect to $c^{\prime} \in C-c$. In simple terms, ${\vec{d}_c \cdot \hat{n}_{c^{\prime}}}$ is a measure of the external associations between clusters~$c$~and~${c^{\prime}}$.  \\

By intuition, as we mentioned earlier at the beginning of section~\ref{sec:intro}, a community is a group of nodes with strong internal associations (cohesion) and weak external associations (separation) with the nodes of other groups.  Therefore, based on this concept of cohesion and separation, for cluster $c$ to be a meaningful group, we propose that ${\vec{d}_c \cdot \hat{n}_c}$ (measure of internal associations within $c$) should be as large as possible and ${\vec{d}_c \cdot \hat{n}_{c^{\prime}}}$ (measure of associations between $c$ and $c^{\prime}$) should be as low as possible $\forall~{c^{\prime}} \in C-c$. Expressing this idea mathematically, 
\begin{IEEEeqnarray}{lll} 
 M_c = \vec{d}_c \cdot \hat{n}_c - \sum_{c^{\prime} \in C-c} \vec{d}_c \cdot \hat{n}_{c^{\prime}}\nonumber\\ \hspace{1.7em}
= \vec{d}_c \cdot (\hat{n}_c - \sum_{c^{\prime} \in C-c}  \hat{n}_{c^{\prime}}),  %
 \label{eq:modulardensity_1}
 \end{IEEEeqnarray} 
where $M_c$ should be as large as possible for cluster $c$ to be a meaningful group. On applying this concept to all the clusters in the network, we introduce the following global measure:
\begin{IEEEeqnarray}{lll} 
 M = \sum_{c \in C} M_c =  \sum_{c \in C} \{\vec{d}_c \cdot (\hat{n}_c - \sum_{c^{\prime} \in C-c}  \hat{n}_{c^{\prime}})\}  \label{eq:modulardensity_2} \\ \hspace{5.7em} 
 =  \sum_{c \in C} \{\vec{d}_c \cdot (2\hat{n}_c - \sum_{c^{\prime} \in C}\hat{n}_{c^{\prime}})\},
 \label{eq:modulardensity_3}
 \end{IEEEeqnarray} 
where we define $M$ as a measure of modularity density and propose that $M$ should be maximized in order to obtain a meaningful community structure of the network. Note that our new measure of modularity density differs from the previous mathematical formulations of modularity density in the literature; we also show in section~\ref{sec:sensitivity studies} how our metric $M$ is superior to these previous versions.

$M$ can be further expressed in terms of $\bold{T}$ by substituting equation~(\ref{eq:degree1}) in (\ref{eq:modulardensity_3}) as,
\begin{IEEEeqnarray}{lll} 
 M  =  \sum_{c \in C} \{\hat{n}_c \cdot \bold{T} \cdot (2\hat{n}_c - \sum_{c^{\prime} \in C}  \hat{n}_{c^{\prime}})\}. \label{eq:modulardensity_4} 
  \end{IEEEeqnarray} 
If the sum of the unit vectors of all the clusters $c^{\prime} \in C$ is represented by $\vec{N}$, \emph{i.e.}
\begin{IEEEeqnarray}{lll} 
~\vec{N} = \sum_{c^{\prime} \in C} \hat{n}_{c^{\prime}}, \label{eq:Nsum}
\end{IEEEeqnarray}
then the expression for modularity density in (\ref{eq:modulardensity_4}) reduces to
\begin{IEEEeqnarray}{lll} 
M =  2\sum_{c \in C} \{\hat{n}_c \cdot \bold{T} \cdot \hat{n}_c\} - \sum_{c \in C}  \hat{n}_c \cdot \bold{T} \cdot \vec{N} \nonumber \\ \hspace{1.35em}
=  2\sum_{c \in C} \{\hat{n}_c \cdot \bold{T} \cdot \hat{n}_c\}  - \vec{N} \cdot \bold{T} \cdot \vec{N}.\label{eq:modulardensity_5} 
\end{IEEEeqnarray}

Alternatively, $M$ can be also expressed using the derivations~(\ref{eq:internaldensity},~\ref{eq:penalty}) in equation~(\ref{eq:modulardensity_2}) as:
\begin{IEEEeqnarray}{lll}
M = \sum_{c \in C}\Bigg\{\frac{\sum_{i,j \in c}T_{ij}}{n_c}  - \sum_{c^{\prime} \in C-c}\Bigg( \frac{\sum_{{i \in c,}{j \in c^{\prime}}}T_{ij}}{\sqrt{n_c n_{c^{\prime}}}}\Bigg)\Bigg\}.~\label{eq:modulardensity_6} 
\end{IEEEeqnarray}

Note that our new measure of modularity density assumes that the graph $G(V,~E)$ is an undirected, unweighted/weighted (no negative edge weight), connected network and that each node in the network belongs to only a single cluster. Therefore, our metric applies only for detecting non-overlapping communities in an undirected, connected network.  
\section{Sensitivity Studies}\label{sec:sensitivity studies}  
Real-world networks are heterogeneous in nature, \emph{i.e.} they comprise communities of varied sizes~\cite{d06,f10}.  Previous studies~\cite{l11} have shown that community detection in heterogeneous networks by means of optimizing a desired metric should be free from bias, \emph{i.e.} the tendency of favoring larger clusters over smaller ones (the resolution limit problem) or the problem of favoring smaller clusters over larger ones. In order to resolve such problems of bias, Li \emph{et al.} (2008)~\cite{l08} and Chen \emph{et al.} (2013)~\cite{c13} have introduced different versions of modularity density. In this section, we mathematically show that our new measure of modularity density, $M$, does not suffer from any such bias, and we further show how our metric $M$ is superior to the previous versions of modularity density.

\subsection{$M$ does not suffer from bias}\label{subsec:bias}
In order to show that our metric $M$ is free from bias, we test our metric on the three following cases, which are more general compared to the example networks used in the literature~\cite{l08,c13}.\\
\paragraph{The metric does not split a random graph or a clique into smaller modules} A connected random graph or a clique is not expected to have communities within its network~\cite{f10}.  In this section, we show that optimizing our metric $M$ does not split a connected random graph or a clique into smaller modules. 

Based on the Erd\"{o}s-R\'{e}nyi model~\cite{e59}, consider a random graph ${G(m, p_m)}$, where $m \geq 3$ is the number of nodes in the network and ${{p_m} \in {[p_{m_{min}},~1]}}$ is the probability of an edge being present between any two nodes in the network. Here ${p_{m_{min}}}$ indicates the minimum edge probability required for the graph to be connected and to form a natural community.  For a graph with $m$ nodes to be connected, at least ${m-1}$ edges are required, while the graph can have at most ${m(m-1)/2}$~edges;  furthermore, as far as forming a natural community is concerned, a natural community of $m$ nodes with the fewest edges is a ring of nodes with $m$ edges\footnote{A natural community is not expected to  have internal communities within~\cite{l11}.  By this definition, a path graph is not a natural community as any connected proper subgraph of a path graph is another path graph, which is loosely connected to the remainder of the original path graph.}. In other words, atleast $m$ edges are required to form a natural community. Therefore, the minimum edge probability for the connected random graph ${G(m, p_m)}$ is:
\begin{IEEEeqnarray}{lll}
p_{m_{min}} = \frac{m}{\frac{m(m-1)}{2}} = \frac{2}{m-1}.  \label{eq:min_probability}
\end{IEEEeqnarray}
Furthermore, for a given $p_m$, the total degree of the random graph is ${p_m\frac{2m (m-1)}{2}}$. Note that when ${p_m = 1}$, ${G(m, p_m)}$ becomes a clique, where every two nodes in the network are connected by an edge. Suppose ${M_{single}}$ represents the modularity density of the random graph when the network is treated as one single community, then using equation~(\ref{eq:modulardensity_6}):
\begin{IEEEeqnarray}{lll}
M_{single} = {p_m} \frac{2m(m-1)}{2m} = {p_m}{(m-1)}. \label{eq:single_1}
\end{IEEEeqnarray}

If the above random graph is split into two clusters $c_1$ and $c_2$ with $m_1$ and $m_2$ nodes, respectively, such that $m_1+m_2 = m$, then using equation~(\ref{eq:modulardensity_6}):
\begin{IEEEeqnarray}{lll}
M_{split} = {{p_m} \frac{2m_1(m_1-1)}{2m1} + {p_m} \frac{2m_2(m_2-1)}{2m_2}} \nonumber \\ \nonumber \\  \hspace{3.9em}
-\> {2}{p_m}\frac{{m_1m_2}}{\sqrt{{m_1m_2}}} \nonumber \\  \nonumber \\  \hspace{3em}
 =  {p_m} {(m_1 + m_2 - 2)} - {2}{p_m}{\sqrt{m_1m_2}}  \nonumber \\  \nonumber \\ \hspace{3em}
 = {p_m}{(m-2)} - {2}{p_m}{\sqrt{m_1m_2}}, \label{eq:split_1}
\end{IEEEeqnarray}
where $M_{split}$ is the modularity density of the above partitioned network. Therefore,
\begin{IEEEeqnarray}{lll}
 M_{single} - M_{split} = (\ref{eq:single_1})-(\ref{eq:split_1})\nonumber \\ \nonumber \\ \hspace{7.4em}
 = p_m(1 + 2{\sqrt{m_1m_2}}) > 0,
\end{IEEEeqnarray}
which means that $M_{single} > M_{split}$.  Likewise, it is not hard to show that when the random graph is split into three or more clusters $M_{single} >  M_{split}$. Therefore, we generalize that optimizing $M$ does not split a random graph or a clique into two or more smaller modules.  This means that our metric does not suffer from the problem of favoring smaller clusters over larger ones.

Note that for the remainder of the section~\ref{subsec:bias}, unless specified, the reader should assume that $M$ is determined using equation~(\ref{eq:modulardensity_6}).\\ 

\begin{figure}[htbp]
\psfrag{m}{$it~works$}
\hspace{-1.5mm}\centerline{\includegraphics[width=2.0in]{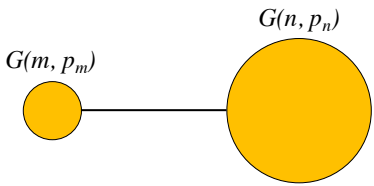}}
\caption{Network of communities of different sizes.} \label{fig:two_random}
\end{figure}

\paragraph{The metric identifies communities of different sizes} Consider a simple heterogeneous network with two natural communities as shown in figure~\ref{fig:two_random}. The communities are of different sizes and are loosely connected by a single edge. Let the two natural communities be random graphs $G(m,~p_m)$ and $G(n,~p_n)$ of the Erd\"{o}s-R\'{e}nyi's model~\cite{e59}.  The parameters $m, n \geq 3$ are the number of nodes, and $p_m \in [p_{m_{min}},~1],~p_n \in [p_{n_{min}}, 1]$ are the edge probabilities of the graphs $G(m,~p_m)$ and $G(n,~p_n)$, respectively. As we defined earlier in section~\ref{subsec:bias}($a$), $p_{m_{min}}$ and $p_{n_{min}}$ are the minimum edge probabilities of their respective networks. Suppose $p_m = p_n = 1$, then the communities in figure~\ref{fig:two_random} are cliques. 
For the case of cliques, previous studies~\cite{c18} have shown that optimizing the Chen \emph{et al.} (2013)'s~\cite{c13} modularity density fails to identify the communities as separate clusters for a certain range of ${\frac{m}{n}}$. In this section, we show that optimizing our metric $M$ successfully identifies the natural communities (cliques or random graphs) as separate clusters for any ${\frac{m}{n}}$.

If $M_{single}$ is the modularity density of the network when the communities in figure~\ref{fig:two_random} are merged into a single cluster, then:
\begin{IEEEeqnarray}{lll}
M_{single} =   \frac{p_m (\frac{2m(m-1)}{2}) +  p_n (\frac{2n(n-1)}{2}) + 2}{m+n}  \nonumber \\ \nonumber \\ \hspace{3.5em}
= \frac{p_m (m-1)m + p_n (n-1)n + 2}{m+n}. \label{eq:single_2}
\end{IEEEeqnarray}
Likewise, when the two communities are split into separate clusters the corresponding modularity density $M_{sep}$ is given as:
\begin{IEEEeqnarray}{lll}
M_{sep} = {{p_m} \frac{2m(m-1)}{2m} + {p_n} \frac{2n(n-1)}{2n}} - \frac{2}{\sqrt{mn}} \nonumber \\ \nonumber \\ \hspace{2.4em}
= {p_m(m-1)} + {p_n(n-1)} - \frac{2}{\sqrt{mn}}. \label{eq:split_2}
\end{IEEEeqnarray}
Determining the difference between $M_{sep}$ and $M_{single}$:
\begin{IEEEeqnarray}{lll}
\Delta M = M_{sep} - M_{single} = (\ref{eq:split_2}) - (\ref{eq:single_2})\nonumber  \\ \nonumber  \\  \hspace{2.2em}
= \Delta I - \frac{2}{\sqrt{mn}} -  \frac{2}{m+n}, \label{eq:delta_0} \\ \nonumber  \\ \nonumber
\hspace{-2.8em}$where~$\hspace{0.2em} \Delta I \hspace{0.25em}= {p_m(m-1)} + {p_n(n-1)} \\ \nonumber  \\ \hspace{3.3em}
- \frac{p_{m} (m-1)m + p_n (n-1)n}{m+n}\nonumber \\ \nonumber  \\ \hspace{2.0em}
=\> \frac{p_{m} (m-1)n + p_n (n-1)m}{m+n}. \label{eq:delta_1}
\end{IEEEeqnarray}
From equation~(\ref{eq:delta_1}), $\Delta I > 0$ as $m,~n \geq 3$ and $p_{m},~p_{n}$ are always positive.~Also, at a given $m$ and $n$, $\Delta I$ is minimum when $p_{m} = p_{m_{min}}$ and $p_{n} = p_{n_{min}}$.  This means:
\begin{IEEEeqnarray}{lll}
\Delta I \geq \Delta I_{min}  = \Delta I \Big|_{(p_m = p_{m_{min}},~p_n = p_{n_{min}})},  \label{eq:delta_2}  \\ \nonumber \\ \nonumber
\Delta I_{min} =  {p_{m_{min}} (m-1)} + {p_{n_{min}}(n-1)}   \\ \nonumber  \\ \hspace{4.2em}
-\> \frac{p_{m_{min}} (m-1)m + p_{n_{min}} (n-1)n}{m+n}.  \label{eq:delta_3}
\end{IEEEeqnarray}
As we derived earlier in equation~(\ref{eq:min_probability}), the minimum edge probabilities for the connected random graphs in figure~\ref{fig:two_random} are $p_{m_{min}} = \frac{2}{m-1}$ and $p_{n_{min}} = \frac{2}{n-1}$.  Substituting these values for edge probabilities in equation~(\ref{eq:delta_3}) gives:
\begin{IEEEeqnarray}{lll}
\Delta I_{min} =  {\frac{2}{m-1} (m-1)} + {\frac{2}{n-1}}(n-1) \nonumber  \\ \nonumber  \\ \nonumber \hspace{4.2em}
-\> \frac{\frac{2}{m-1} (m-1)m + \frac{2}{n-1} (n-1)n}{m+n}  \\ \nonumber  \\  \hspace{3.0em}
=4  - \frac{2(m+n)}{m+n} = 2.  \label{eq:delta_4}
\end{IEEEeqnarray}
Using the result (\ref{eq:delta_4}) and the inequality (\ref{eq:delta_2}) in equation~(\ref{eq:delta_0}), the difference between $M_{sep}$ and $M_{single}$ is:
\begin{IEEEeqnarray}{lll}
\Delta M \geq \Delta I_{min} - \frac{2}{\sqrt{mn}} - \frac{2}{m+n}\nonumber \\ \nonumber  \\ \nonumber \hspace{2.2em}
= 2  - \frac{2}{\sqrt{mn}} - \frac{2}{m+n}. 
\end{IEEEeqnarray}
Since $m, n \geq 3$,
\begin{IEEEeqnarray}{lll}
 \hspace{-1.7em} \Delta M \geq 2  - \frac{2}{\sqrt{3 \times 3}} - \frac{2}{3+3}  \nonumber  \\ \nonumber  \\ \hspace{0.5em}
= 1,
\end{IEEEeqnarray}
which implies that $\Delta M > 0$, $i.e.$ $M_{sep} > M_{single}$ and indicates that optimizing our metric $M$ identifies the two natural communities in figure~\ref{fig:two_random} as separate clusters for all $\frac{m}{n}$.  This result is promising and suggests that our metric does not suffer from the tendency of favoring larger clusters over smaller ones. \newline

\begin{figure}[htbp]
\hspace{0.0mm}\centerline{\includegraphics[width=2.9in]{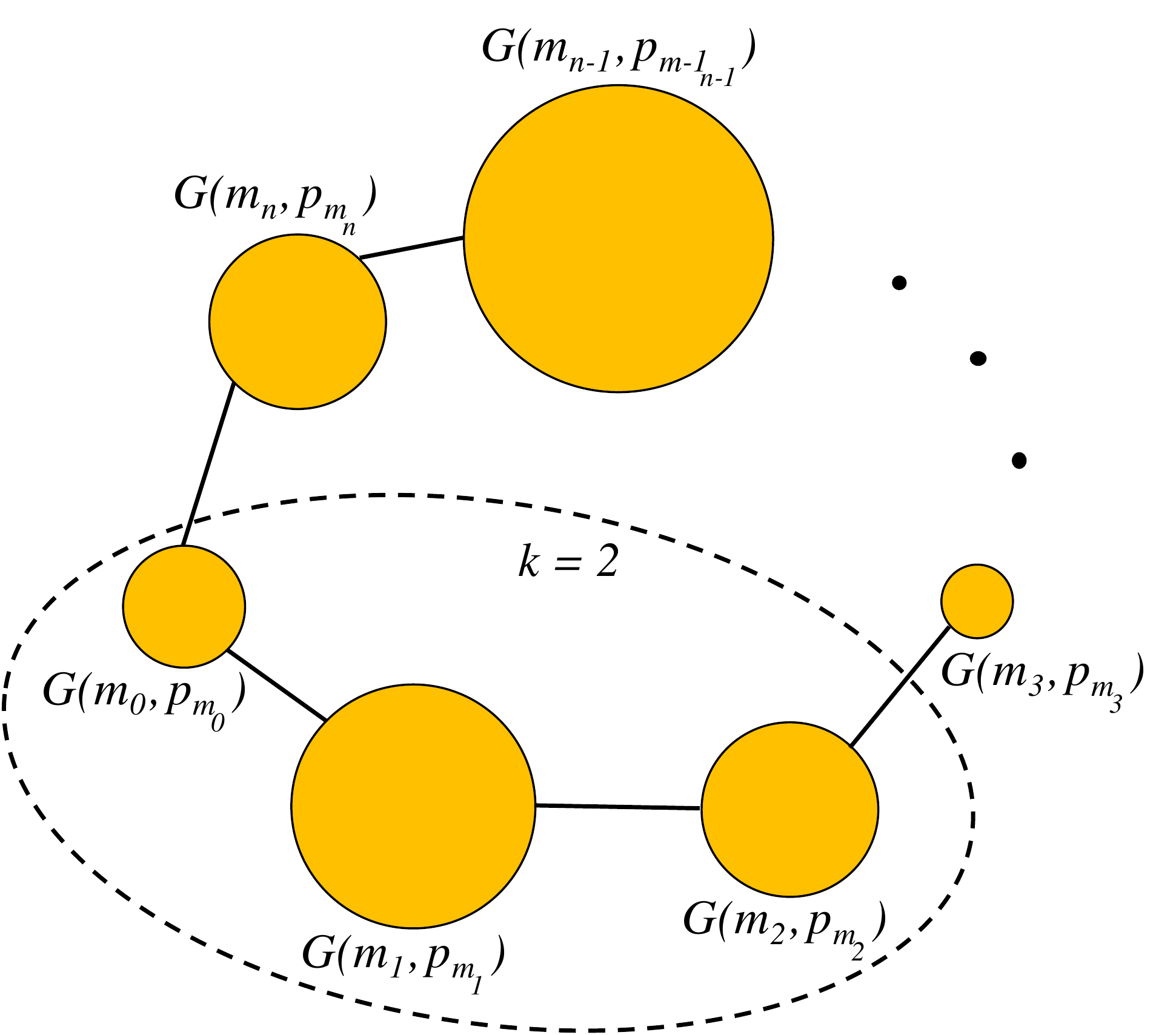}}
\caption{Heterogeneous modular network.}\label{fig:modular}
\end{figure}

\paragraph{The metric detects communities in heterogeneous modular networks}  Unlike the example network used in figure~\ref{fig:two_random}, real networks often comprise more than two clusters. Owing to the nature of real networks, we test the performance of our metric by employing a generic heterogeneous network shown in figure~\ref{fig:modular}. This network comprises a ring of $n+1 \geq 3$ natural communities of varied sizes, where the communities adjacent to each other are connected by a single edge. Once again, let these $n+1$ natural communities be random graphs $G(m_0, p_{m_0}),~G(m_1, p_{m_1}),...,~G(m_n, p_{m_n})$ of the Erd\"{o}s-R\'{e}nyi model~\cite{e59}.  For each $i \in \{0, 1, 2,..., n\}$,  $m_i \geq 3$ is the number of nodes, $p_{m_i} \in [p_{{m_i}_{min}},~1]$ is the edge probability and $p_{{m_i}_{min}}$ is the minimum edge probability of $G(m_i, p_{m_i})$.  
\begin{figure}[htbp]
\hspace{0.0mm}\centerline{\includegraphics[width=3.0in]{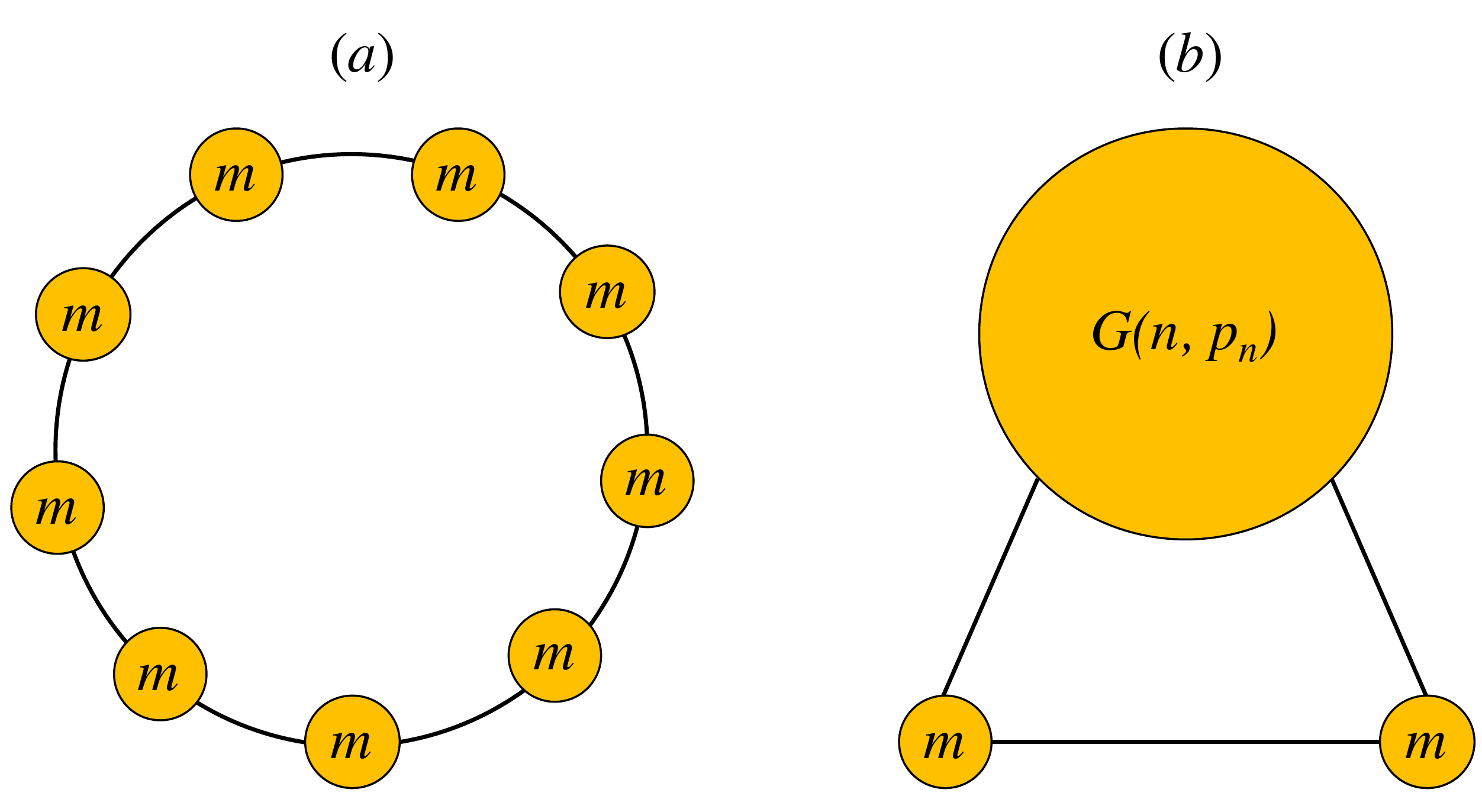}}
\caption{(\textit{a}) Ring of cliques. (\textit{b}) Network with two cliques and a random subgraph.}\label{fig:examples}
\end{figure}
For any $i \in \{0, 1, 2,...,n\}$ if $p_{m_i} = 1$, then the corresponding network $G(m_i, p_{m_i})$ becomes a clique. Note that the sample networks presented in figure~\ref{fig:examples} are all specific examples of the generic heterogeneous network in figure~\ref{fig:modular}. In the current section, we show that optimizing our metric $M$ successfully identifies all the true communities of this generic heterogeneous network as separate clusters.

If $M_{sep}$ represents the modularity density of the heterogeneous network when all the natural communities in figure~\ref{fig:modular} are identified as separate clusters, then using equation~(\ref{eq:modulardensity_6}):
\begin{IEEEeqnarray}{lll}
M_{sep} = \sum_{i=0}^{n} \frac{p_{m_i} 2{m_i}({m_i}-1)}{2{m_i}}  -  \sum_{i=0}^{n-1}\frac{2}{\sqrt{{m_i}{m_{i+1}}}} - \frac{2}{\sqrt{{m_0}{m_n}}} \nonumber \\ \nonumber \\ \hspace{2.4em}
= \sum_{i=0}^{n} {p_{m_i}({m_i}-1)}  -  \sum_{i=0}^{n-1}\frac{2}{\sqrt{{m_i}{m_{i+1}}}} - \frac{2}{\sqrt{{m_0}{m_n}}}.\label{eq:split_3} 
\end{IEEEeqnarray}
Let us now consider merging some of these natural communities into a larger group. Say for any $k \in \{1, 2,...{n-1}\}$ we merge $k+1$ communities, \emph{i.e.} we merge $G(m_0, p_{m_0}),~G(m_1, p_{m_1}),...,~G(m_k, p_{m_k})$ into a single cluster. A sample illustration of merging is shown in figure~\ref{fig:modular}, which shows that when $k = 2$ we merge the three communities $G(m_0, p_{m_0}),~G(m_1, p_{m_1}) $and$~G(m_2, p_{m_2})$ into a single group.  If $M_{merge}^{k}$ denotes the modularity density of the network when the above defined $k + 1 \leq n$ communities are merged and the remaining $n-k$ communities are treated as separate clusters, then from equation~(\ref{eq:modulardensity_6}):
\begin{IEEEeqnarray}{lll} 
M_{merge}^{k} =  \frac{2k + \sum_{i=0}^{k} p_{m_i} (\frac{2m_i(m_i-1)}{2})}{\sum_{i=0}^{k} m_i} \nonumber \\ \nonumber  \\ \nonumber \hspace{4.8em}
+\> \sum_{i=k+1}^{n}\frac{p_{m_i} 2{m_i}({m_i}-1)}{2{m_i}} -\sum_{i=k+1}^{n-1}\frac{2}{\sqrt{m_i m_{i+1}}} \\ \nonumber \hspace{4.8em}
-\> \frac{2}{\sqrt{m_{k+1} \sum_{i=0}^{k}m_i}}- \frac{2}{\sqrt{m_{n} \sum_{i=0}^{k}m_i}}\\ \nonumber \\ \nonumber \hspace{3.5em}
=  \frac{2k + \sum_{i=0}^{k} p_{m_i} {(m_i-1)m_i}}{\sum_{i=0}^{k} m_i}  \\ \nonumber  \\ \nonumber  \hspace{4.8em}
+\> \sum_{i=k+1}^{n}{p_{m_i}({m_i}-1)}  -\sum_{i=k+1}^{n-1}\frac{2}{\sqrt{m_i m_{i+1}}} \\ \nonumber  \\  \hspace{4.8em}
-\> \frac{2}{\sqrt{\sum_{i=0}^{k}m_i}}\Big( \frac{1}{\sqrt{m_{k+1}}}+ \frac{1}{\sqrt{m_{n}}} \Big).\label{eq:mergek}
\end{IEEEeqnarray}
Taking the difference between $M_{sep}$ and $M_{merge}^{k}$ gives:
\begin{IEEEeqnarray}{lll}
\Delta M^{k} = M_{sep} - M_{merge}^{k} =  (\ref{eq:split_3}) - (\ref{eq:mergek})\nonumber\\ \nonumber \\ \nonumber  \hspace{2.65em}
=  \Delta I^{k}  - \frac{2k}{\sum_{i=0}^{k} m_i} - \sum_{i=0}^{k}\frac{2}{\sqrt{{m_i}{m_{i+1}}}} - \frac{2}{\sqrt{{m_0}{m_n}}} \\ \nonumber  \\  \hspace{4.0em}
+\> \frac{2}{\sqrt{\sum_{i=0}^{k}m_i}}\Big( \frac{1}{\sqrt{m_{k+1}}}+ \frac{1}{\sqrt{m_{n}}} \Big)  , \label{eq:deltamk_1}
\end{IEEEeqnarray}
where the expression for $\Delta I^k$ is: 
\begin{IEEEeqnarray}{lll}
\Delta I^{k} =  \sum_{i=0}^{k} {p_{m_i}({m_i}-1)} - \frac{\sum_{i=0}^{k} p_{m_i} {(m_i-1)m_i}}{\sum_{i=0}^{k} m_i}\label{eq:deltak}  \\ \nonumber \\ \hspace{2.0em}
=  \frac{\sum_{i=0}^{k} \Big(p_{m_i} {(m_i-1) \sum_{j=0}^{k} m_{j \neq i}} \Big)}{\sum_{i=0}^{k} m_i}. \label{eq:deltak2}
\end{IEEEeqnarray}
\\
Note that $\forall~i \in \{0, 1,2,...,k\},~m_i \geq 3$ and ${p_{m_i} \geq p_{m_{i_{min}}}> 0}$. Therefore, from equation~(\ref{eq:deltak2}), $\Delta I^k > 0$.  Furthermore, for a given ${m_i}$, $\Delta I^k$ is minimum when $p_{m_i} = p_{m_{i_{min}}}$ for all the $k+1$ communities. In mathematical terms:
\begin{IEEEeqnarray}{lll}
\Delta I^{k} \geq \Delta I_{min}^{k} = \Delta I^{k} \Big |_{(\forall i \in \{0, 1,...,k\},~p_{m_i} = p_{{m_i}_{min}})} \label{eq:deltak_min}
\end{IEEEeqnarray}\\
From the previously derived expression (\ref{eq:min_probability}) for the minimum edge probability, we get $p_{m_{i_{min}}} = \frac{2}{m_i -1}$.  Using these values for the edge probabilities $\forall~i \in \{0, 1,...,k\}$ in equation~(\ref{eq:deltak}), we obtain the minimum $\Delta I^{k}$ as:
\begin{IEEEeqnarray}{lll}
\Delta I_{min}^{k} = \sum_{i=0}^{k} {\frac{2}{m_i-1}({m_i}-1)} - \frac{\sum_{i=0}^{k} \frac{2}{m_i-1} {(m_i-1)m_i}}{\sum_{i=0}^{k} m_i} \nonumber \\ \nonumber \\  \hspace{3.0em}
= 2(k+1)  - \frac{{2}\sum_{i=0}^{k} {m_i}}{\sum_{i=0}^{k} m_i} = 2k. 
\end{IEEEeqnarray}\\
Applying this result for the minimum value of $\Delta I^k$  in equation~(\ref{eq:deltamk_1}) gives the inequality:
\begin{IEEEeqnarray}{lll}
\Delta M^{k} \geq  \Delta I_{min}^{k}  - \frac{2k}{\sum_{i=0}^{k} m_i} - \sum_{i=0}^{k}\frac{2}{\sqrt{{m_i}{m_{i+1}}}} - \frac{2}{\sqrt{{m_0}{m_n}}} \nonumber  \\ \nonumber  \\ \nonumber  \hspace{3.8em}
+\>  \frac{2}{\sqrt{\sum_{i=0}^{k}m_i}}\Big( \frac{1}{\sqrt{m_{k+1}}}+ \frac{1}{\sqrt{m_{n}}} \Big)  \\ \nonumber  \\ \nonumber \hspace{2.65em}
= 2k - \frac{2k}{\sum_{i=0}^{k} m_i} - \sum_{i=0}^{k}\frac{2}{\sqrt{{m_i}{m_{i+1}}}} - \frac{2}{\sqrt{{m_0}{m_n}}}\\ \nonumber  \\  \hspace{3.8em}
+\>   \frac{2}{\sqrt{\sum_{i=0}^{k}m_i}}\Big( \frac{1}{\sqrt{m_{k+1}}}+ \frac{1}{\sqrt{m_{n}}} \Big). \label{eq:deltamk_2} 
\end{IEEEeqnarray}
\newline 
Given that $m_0, m_n \geq 3$ and $m_i \geq 3$ for all the $k+1$ communities in equation~(\ref{eq:deltamk_2}), we deduce the following: 
\begin{IEEEeqnarray}{lll}
\nonumber  \\ \nonumber  \\ \nonumber
\Delta M^{k} \geq 2k - \frac{2k}{\sum_{i=0}^{k} 3} - \sum_{i=0}^{k}\frac{2}{\sqrt{{3}\times{3}}}-\frac{2}{\sqrt{3 \times 3}} \nonumber  \\ \nonumber  \\ \nonumber  \hspace{3.8em}
+\> \frac{2}{\sqrt{\sum_{i=0}^{k}m_i}}\Big( \frac{1}{\sqrt{m_{k+1}}}+ \frac{1}{\sqrt{m_{n}}} \Big)\\ \nonumber  \\ \nonumber  \hspace{2.65em}
= 2k - \frac{2k}{3(k+1)} - \frac{2}{3}(k+1)-\frac{2}{3} \\ \nonumber  \\ \nonumber  \hspace{3.8em}
+\> \frac{2}{\sqrt{\sum_{i=0}^{k}m_i}}\Big( \frac{1}{\sqrt{m_{k+1}}}+ \frac{1}{\sqrt{m_{n}}} \Big)\\ \nonumber  \\ \nonumber \hspace{2.65em}
=  \frac{4(k-1)}{3} - \frac{2k}{3(k+1)}  \\ \nonumber \\ \nonumber \hspace{4.0em}
+\> \frac{2}{\sqrt{\sum_{i=0}^{k}m_i}}\Big( \frac{1}{\sqrt{m_{k+1}}}+ \frac{1}{\sqrt{m_{n}}}\Big) \\ \nonumber \\ \nonumber \hspace{2.65em}
= \frac{4(k-2)^2+14(k-2)+8}{3(k+1)} \\ \nonumber \\  \hspace{4.0em}
+\> \frac{2}{\sqrt{\sum_{i=0}^{k}m_i}}\Big( \frac{1}{\sqrt{m_{k+1}}}+ \frac{1}{\sqrt{m_{n}}}\Big) \nonumber \\ \nonumber \\ \hspace{2.65em} 
\implies \Delta M^{k} > 0~~$~if~$  k \in \{2, 3,...,{n-1}\} \label{eq:deltamk_3} ,
\end{IEEEeqnarray}
which means that $M_{sep} > M_{merge}^{k}$ for all the integer values of $k$ in the interval $2 \leq k \leq {n-1}$.  This is a positive result. However, based on the range of values we defined earlier for $k$, the above result does not completely prove that $M_{sep} > M_{merge}^{k}$ $\forall~k \in \{1, 2,...,{n-1}\}$.  We still have to show that $M_{sep} > M_{merge}^{k}$ when $k=1$, which indicates the case of merging the two communities $G(m_0, p_{m_0})$ and   
$G(m_1, p_{m_1})$ into one cluster while retaining the remaining $n-1$ communities as separate clusters. Therefore, by reusing~(\ref{eq:deltamk_2}), we obtain $\Delta M^{k}$ for $k = 1$ as:
\begin{IEEEeqnarray}{lll}
\Delta M^1 \geq 2 - \frac{2}{m_0 + m_1} - \frac{2}{\sqrt{{m_0}{m_{1}}}} - \frac{2}{\sqrt{{m_1}{m_{2}}}} - \frac{2}{\sqrt{{m_0}{m_n}}}\nonumber \\ \nonumber  \\  \hspace{3.8em}
+\>   \frac{2}{\sqrt{m_0 + m_1}}\Big( \frac{1}{\sqrt{m_{2}}}+ \frac{1}{\sqrt{m_{n}}} \Big). 
\end{IEEEeqnarray}
On rearranging the terms on the right-hand side (RHS) of the above inequality, we have: 
\begin{IEEEeqnarray}{lll}
\Delta M^1 \geq 2 - \frac{2}{m_0 + m_1} - \frac{2}{\sqrt{{m_0}{m_{1}}}} \nonumber \\ \nonumber  \\  \hspace{3.8em}
-\> \frac{2}{\sqrt{m_2}}\Big(\underbrace{\frac{1}{\sqrt{m_1}} - \frac{1}{\sqrt{m_0 + m_1}}}_{\text{$a$}}\Big) \nonumber \\ \nonumber  \\  \hspace{3.8em}
-\> \frac{2}{\sqrt{m_n}}{\Big(\underbrace{\frac{1}{\sqrt{m_0}} - \frac{1}{\sqrt{m_0 + m_1}}}_{\text{$b$}}\Big)}. \label{eq:deltam1_1}
\end{IEEEeqnarray}
As $\sqrt{m_1} < \sqrt{m_0 + m_1}$ and $\sqrt{m_0} < \sqrt{m_0 + m_1}$, the terms $a$ and $b$ as indicated in~(\ref{eq:deltam1_1}) are always positive.  Given that $m_2, m_n \geq 3$ along with $a, b > 0$, we derive from (\ref{eq:deltam1_1}):
\begin{IEEEeqnarray}{lll}
\Delta M^1 \geq 2 - \frac{2}{m_0 + m_1} - \frac{2}{\sqrt{{m_0}{m_{1}}}} \nonumber \\ \nonumber  \\  \hspace{3.8em}
-\> \frac{2}{\sqrt{3}}\Big(\frac{1}{\sqrt{m_1}} - \frac{1}{\sqrt{m_0 + m_1}}\Big) \nonumber \\ \nonumber  \\  \hspace{3.8em}
-\> \frac{2}{\sqrt{3}}{\Big(\frac{1}{\sqrt{m_0}} - \frac{1}{\sqrt{m_0 + m_1}}\Big)} \label{eq:deltam1_2}
\end{IEEEeqnarray}
Once again, on rearranging the RHS terms of the inequality (\ref{eq:deltam1_2}), we obtain:
\begin{IEEEeqnarray}{lll}
\Delta M^1 \geq  2 - \frac{2}{\sqrt{m_0 m_1}} - \frac{2}{\sqrt{3m_1}} - \frac{2}{\sqrt{3 m_0}} \nonumber \\ \nonumber \\ \nonumber \hspace{3.7em}
+\> \frac{2}{\sqrt{m_0 + m_1}}\Big(\underbrace{\frac{2}{\sqrt{3}} -\frac{1}{\sqrt{m_0+m_1}}}_{\text{$c$}}\Big). \label{eq:deltam1_3}
\end{IEEEeqnarray}
Note that the denoted expression for $c$ in ~(\ref{eq:deltam1_3}) is positive as $m_0,~m_1 \geq 3$. Furthermore, for this range of $m_0$ and $m_1$, we infer:
\begin{IEEEeqnarray}{lll}
\Delta M^1 \geq  2 - \frac{2}{\sqrt{3 \times 3}} - \frac{2}{\sqrt{3 \times 3}} - \frac{2}{\sqrt{3 \times 3}} \nonumber \\ \nonumber \\ \nonumber \hspace{3.7em}
+\> \frac{2}{\sqrt{m_0 + m_1}}\Big(\frac{2}{\sqrt{3}} -\frac{1}{\sqrt{m_0+m_1}}\Big)  \nonumber \\ \nonumber \\ \nonumber \hspace{3.5em}
= 2 - \frac{2}{3} \times 3 + \frac{2}{\sqrt{m_0 + m_1}}\Big(\frac{2}{\sqrt{3}} -\frac{1}{\sqrt{m_0+m_1}}\Big)  \nonumber \\ \nonumber \\  \hspace{3.5em}
= \frac{2}{\sqrt{m_0 + m_1}}\Big(\frac{2}{\sqrt{3}} -\frac{1}{\sqrt{m_0+m_1}}\Big) > 0  \nonumber \\ \nonumber \\  \hspace{3.2em}
\implies \Delta M^1 > 0, \label{eq:deltam1_4}
\end{IEEEeqnarray}
which means that $M_{sep} > M_{merge}^{k}$ for $k = 1$. Hence, from the results~(\ref{eq:deltamk_3}, \ref{eq:deltam1_4}), $M_{sep} > M_{merge}^k$ $~\forall~k \in \{1, 2,...,{n-1}\}$, which defines cases of merging $k+1 \leq n$ communities.  Finally, if we consider merging all the $n+1$ communities in figure~\ref{fig:modular} into one big cluster, the corresponding modularity density $M_{merge}^n$ is: 
\begin{IEEEeqnarray}{lll}
M_{merge}^n =  \frac{2n + \sum_{i=0}^{n} p_{m_i} (\frac{2m_i(m_i-1)}{2})}{\sum_{i=0}^{n} m_i}.
\end{IEEEeqnarray}
Along the lines of the proof we provided for (\ref{eq:deltamk_3}) \& (\ref{eq:deltam1_4}), it is not hard to show that
\begin{IEEEeqnarray}{lll}
M_{sep} > M_{merge}^n. \label{eq:deltam_all}
\end{IEEEeqnarray}
These results (\ref{eq:deltamk_3}), (\ref{eq:deltam1_4}) \& (\ref{eq:deltam_all}) point out that optimizing $M$ successfully identifies all the natural communities in the heterogeneous network (figure~\ref{fig:modular}) as separate clusters.  This means that the metric $M$ does not suffer from the resolution limit problem.\\

To summarize the results of section~\ref{subsec:bias}, the subsection \ref{subsec:bias}($a$) shows that our metric $M$ is free from the problem of favoring smaller clusters over larger ones, whereas the subsections \ref{subsec:bias}($b \& c$) show that $M$ does not suffer from the tendency of favoring larger clusters over smaller ones. Therefore, from \ref{subsec:bias}($a,~b,~\&~c$), our metric modularity density $M$ is free from bias.\\

\subsection{$M$ performs better than the previous versions of modularity density}\label{sec:comparse}
Having proved that our metric $M$ is free from the two problems of bias, in the current section we compare the performance of our metric $M$ with that of Li \emph{et al.} (2008)'s~\cite{l08} and Chen \emph{et al.} (2013)'s~\cite{c13} modularity densities and show how $M$ is better than these previous versions. \\
 
\begin{figure}[htbp]
\hspace{0.5mm}\centerline{\includegraphics[width=1.8in]{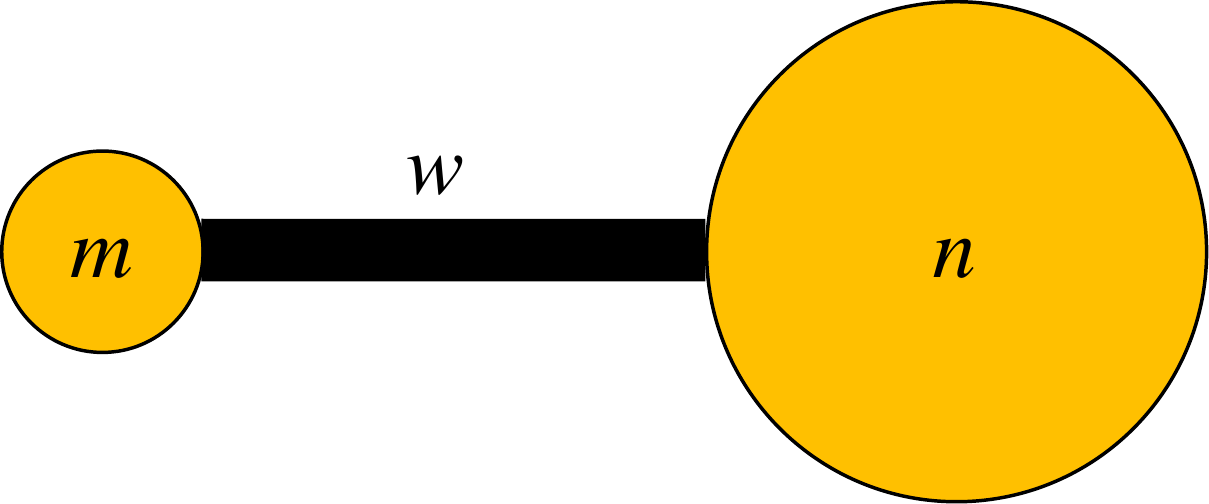}}
\caption{Network of weakly separated communities of different sizes.} \label{fig:twoclique}
\end{figure}

\paragraph{M performs better than the known versions of modularity density in detecting weakly separated communities in heterogeneous networks} Consider a simple unweighted heterogeneous network shown in figure~\ref{fig:twoclique}.  The network comprises two cliques of different sizes.  As in figure~\ref{fig:twoclique}, $m \geq 3$ and $n \geq 3$ represent the number of nodes of the left and right cliques of the heterogeneous network.  While this network looks similar to the one shown earlier in figure~\ref{fig:two_random}, the cliques in figure~\ref{fig:twoclique} are connected by multiple edges.  Let $w$ indicate the total number of edges connecting the two cliques.  We show that optimizing $M$ successfully detects the two cliques as separate clusters for a wider range of $w$ compared to that of Li \emph{et al.} (2008)'s~\cite{l08} modularity density.  Where Chen \emph{et al.} (2013)'s~\cite{c13} modularity density is concerned, previous studies~\cite{c18} have shown that this metric suffers from the resolution limit problem even when the cliques in figure~\ref{fig:twoclique} are loosely connected by a single edge.

Based on our metric, let $M_{merge}$ denote the modularity density of the network when the two cliques in figure~\ref{fig:twoclique} are merged into a single cluster.  Using equation~(\ref{eq:modulardensity_6}), we obtain:  
\begin{IEEEeqnarray}{lll}
M_{merge} = \frac{\frac{2m(m-1)}{2} +  \frac{2n(n-1)}{2} + 2w}{m+n} \nonumber \\ \nonumber \\ \hspace{4.0em}
=  \frac{{m(m-1)} +  {n(n-1)} + 2w}{m+n} \label{eq:merge}
\end{IEEEeqnarray}
When the two cliques are considered as separate clusters, then the corresponding modularity density $M_{sep}$ from equation~(\ref{eq:modulardensity_6}) is:
\begin{IEEEeqnarray}{lll}
M_{sep} = (m-1) + (n-1) - \frac{2w}{\sqrt{mn}} \label{eq:split}. 
\end{IEEEeqnarray}
The difference between $M_{sep}$ and $M_{merge}$ is: 
\begin{IEEEeqnarray}{lll}
\Delta M = M_{sep} - M_{merge} = (\ref{eq:split}) - (\ref{eq:merge})\nonumber \\ \nonumber \\ 
\Delta M =  \frac{{n(m-1)} +  {m(n-1)}}{m+n} - \frac{2w}{m+n} - \frac{2w}{\sqrt{mn}}. \label{eq:deltaM}
\end{IEEEeqnarray} 
From equation~(\ref{eq:deltaM}), $\Delta M > 0$ only if
\begin{IEEEeqnarray}{lll}
w < w_M = \frac{ {{n(m-1)} +  {m(n-1)}}}{2(1+ \frac{m+n}{\sqrt{mn}})}. \label{eq:w_M}
\end{IEEEeqnarray}
This indicates that optimizing $M$ classifies the two cliques in figure~\ref{fig:twoclique} as separate clusters only when $w$ is less than the limiting value $w_M$ given in (\ref{eq:w_M}). \\

\begin{figure}[htbp]
\psfrag{y}{$x$}
\hspace{0.5mm}\centerline{\includegraphics[width=2.2in]{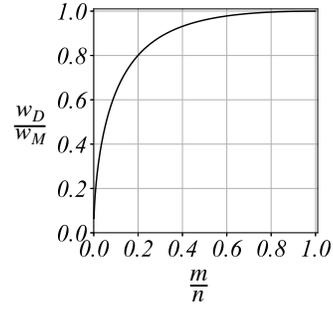}}
\vspace{-7.5mm}\caption{Relation between the limiting values ${w_D}$ and ${w_M}$}~\label{fig:threshold}\vspace{-7.5mm}
\end{figure}

In order to determine the limiting value of $w$ for the case of Li \emph{et al.} (2008)'s~\cite{l08} modularity densitity, we repeat the above procedure (\ref{eq:merge}-\ref{eq:w_M}).  Let $D_{merge}$ and $D_{sep}$ represent Li \emph{et al.} (2008)'s~\cite{l08} modularity density counterparts of $M_{merge}$ and $M_{sep}$, respectively.  Following the definition of Li \emph{et al.} (2008)'s~\cite{l08} modularity density, we obtain $D_{merge}$ and $D_{sep}$ as:
\begin{IEEEeqnarray}{lll} 
D_{merge} =\frac{{m(m-1)} +  {n(n-1)} + 2w}{m+n} \label{eq:merged}
\\ \nonumber \\ 
\&~D_{sep} =  (m-1) + (n-1) - \frac{w}{m} - \frac{w}{n}, \label{eq:splitd}
\end{IEEEeqnarray}
respectively.  From equations~(\ref{eq:merged}~\&~\ref{eq:splitd}), the difference between $D_{sep}$ and $D_{merge}$ is:
\begin{IEEEeqnarray}{lll}
\Delta D = D_{sep} - D_{merge} = (\ref{eq:splitd}) - (\ref{eq:merged}) \\ \nonumber \\ \nonumber
\Delta D =  \frac{{n(m-1)} +  {m(n-1)}}{m+n} - \frac{2w}{m+n} - \frac{w(m+n)}{{mn}}. 
\end{IEEEeqnarray}
Note that $\Delta D >0$ only if
\begin{IEEEeqnarray}{lll}
w < w_D = \frac{ {{n(m-1)} +  {m(n-1)}}}{2(1+ \frac{(m+n)^2}{{2mn}})}. \label{eq:w_D}
\end{IEEEeqnarray}
This means that optimizing Li \emph{et al.} (2008)'s~\cite{l08} modularity density identifies the two cliques in figure~\ref{fig:twoclique} as separate clusters as long as $w$ is less than the limiting value $w_D$ in (\ref{eq:w_D}).  To compare the limiting values $w_D$ and $w_M$, we determine:
\begin{IEEEeqnarray}{lll}
\frac{w_M}{w_D} - 1 = \frac{(\ref{eq:w_M})}{(\ref{eq:w_D})} - 1=  \frac{\frac{(m+n)^2}{{2mn}} -  \frac{m+n}{\sqrt{mn}}}{1+ \frac{m+n}{\sqrt{mn}}} \label{eq:ratioW}  \\ \nonumber \\ \nonumber \hspace{3.8em}
= \Big(\frac{m+n}{2\sqrt{mn}}\Big) \Big(\frac{m+n-2\sqrt{mn}}{1+ \frac{m+n}{\sqrt{mn}}} \Big)  \\ \nonumber \\ \nonumber \hspace{3.8em}
= \Big(\frac{m+n}{2\sqrt{mn}}\Big)  \frac{(\sqrt{m}-\sqrt{n})^2}{1+ \frac{m+n}{\sqrt{mn}}}  \geq 0 \\ \nonumber \\ \hspace{15.0em}
\implies \frac{w_M}{w_D} \geq 1. \label{eq:compareW} 
\end{IEEEeqnarray}
Given the number of nodes $m, n \geq 3$, the above result~(\ref{eq:compareW}) indicates that $w_M$ is always greater than or equal to $w_D$.  A sample illustration of this result is also presented in~figure~\ref{fig:threshold}, which demonstrates the relationship between $w_M$ and $w_D$ based on~equation~(\ref{eq:ratioW}).    
As shown in figure~\ref{fig:threshold}, the limiting value $w_M$ equals $w_D$ only when the network is homogeneous, \emph{i.e.} $m=n$.  In the case of heterogeneous networks, \emph{i.e.} $m \neq n$, the limiting value $w_D$ is always less than $w_M$.  Additionally, figure~\ref{fig:threshold} also depicts that larger the heterogeneity between the communities, greater is the difference between $w_D$ and $w_M$.\\

In conclusion, the results  ($\ref{eq:compareW}$) and figure~\ref{fig:threshold} prove that optimizing $M$ identifies the two communities of the heterogeneous network (figure~\ref{fig:twoclique}) as separate clusters for a wider range of $w$ compared to that of Li \emph{et al.} (2008)'s~\cite{l08} modularity density.  In simple words, our results show that $M$ performs better than the previous versions of modularity density in identifying weakly separated communities in heterogeneous networks.\\

\paragraph{Computing $M$ is on par or faster than the previous versions of modularity density}
To determine the computational complexity of $M$, consider equation~(\ref{eq:modulardensity_5}), which expresses $M$ in terms of ${\hat{n}_c}$,~${\bold{T}}$~and~${\vec{N}}$.  If the adjacency matrix $\bold{T}$ is sparse for a connected network, then the complexity of computing ${\vec{N} \cdot \bold{T}}$ is $O(|E|)$ and the cost of computing ${\vec{N} \cdot \bold{T} \cdot \vec{N}}$ is ${O(|E| + |V|) \approx O(|E|)}$.  Note that $|E|$ and $|V|$ are the number of edges and the number of nodes, respectively, of a given network.  Unlike ${\vec{N} \cdot \bold{T} \cdot \vec{N}}$, the expression ${\hat{n}_c \cdot \bold{T} \cdot \hat{n}_c}$ in equation~(\ref{eq:modulardensity_5}) can be computed at a lower cost as the latter represents the mean internal degree of cluster $c$.  This means the cost of computing ${\hat{n}_c \cdot \bold{T} \cdot \hat{n}_c}$ can be achieved at $O(|E_c|)$, where $|E_c|$ is the number of internal edges of cluster $c$.  Furthermore, the total time complexity of ${\sum_{c \in C} {\hat{n}_c \cdot \bold{T} \cdot \hat{n}_c}}$ is $\sum_{c \in C}O(|E_c|) \approx O(|E|)$.  On combining the complexities of  ${\vec{N} \cdot \bold{T} \cdot \vec{N}}$ and ${\sum_{c \in C} {\hat{n}_c \cdot \bold{T} \cdot \hat{n}_c}}$, the total running time of $M$ in~(\ref{eq:modulardensity_5}) is $O(|E|)$.  Therefore, the cost of computing $M$ is same as the computational cost of Li~\emph{et al.} (2008)'s~\cite{l08} modularity density and faster than that of Chen \emph{et al.} (2013)'s~\cite{c13} metric as the latter has an additional complexity of $O(|C|^2)$ due to the split penalty term~\cite{c14}.  Note that when $\bold{T}$ is dense, $O(|E|)$ approaches $O(|V|^2)$.  Thus, the worst-case running time of $M$ is $O(|V|^2)$.\\

\section{Characteristics of Graph Partitioning using the New Metric}\label{sec:normcut} 
{In this section, we derive some interesting theoretical results that reveal how partitioning a network by maximization of the modularity density $M$ is related to partitioning by minimization of the normalized cut criterion \cite{s00} when subjected to additional constraints.  To derive these results, we consider bi-partitioning an existing cluster of a network and find a partition that maximizes the metric $M$}.\\

Let an undirected network $G(V, E)$, with a set of nodes $V$ and a set of edges $E$ with no negative edge-weights, be partitioned into a set of clusters $C$.  Using equation~(\ref{eq:modulardensity_5}), the modularity density of this network is expressed as: 
\begin{IEEEeqnarray}{lll}
M = \sum_{i \in C} {2\hat{n}_i} \cdot \bold{T}\cdot{\hat{n}_i} - \vec{N}\cdot\bold{T}\cdot\vec{N}. 
\end{IEEEeqnarray}
Alternatively, the above expression of $M$ can be written as:  
\begin{IEEEeqnarray}{lll}
M = 2\hat{n}_c\cdot\bold{T}\cdot\hat{n}_c +  \sum_{i \in C-c} {2\hat{n_i}}\cdot\bold{T}\cdot{\hat{n}_i} - \vec{N}\cdot\bold{T}\cdot\vec{N},~\label{eq:re-mod1}
\end{IEEEeqnarray}
where $\hat{n}_c \in \R^{|V|}$ is a unit vector representing the cluster $c \in C$ as in equation~(\ref{eq:unitvector});  the tensor $\bold{T}$ and the vector $\vec{N}$ are same as what we defined earlier in section~\ref{sec:definition}.  If we now consider bi-partitioning the cluster $c$ into two groups $a$ and $b$, such that:
\begin{IEEEeqnarray}{lll}
n_c = n_a + n_b, \label{eq:bisect}
\end{IEEEeqnarray}
where $n_a$, $n_b$ and $n_c$ are the number of nodes of the clusters $a$, $b$ and $c$, respectively, then the corresponding modularity density $M +  \delta M$ of the network is:
\begin{IEEEeqnarray}{lll}
 M+ \delta M = 2\hat{n}_a\cdot\bold{T}\cdot\hat{n}_a + 2\hat{n}_b\cdot\bold{T}\cdot\hat{n}_b + \sum_{i \in {C - c}} {2\hat{n}_i}\cdot\bold{T}\cdot{\hat{n}_i} \nonumber\\ \hspace{5.4em}
- {(\vec{N} + \delta \vec{N})}\cdot\bold{T}\cdot(\vec{N} + \delta \vec{N}),~\label{eq:re-mod2} \vspace{0.7em} \\
\hspace{-0.4em}$with$~\delta \vec{N} = \hat{n}_a + \hat{n}_b - \hat{n}_c.~\label{eq:deltaN}
\end{IEEEeqnarray}
The change in modularity density as a result of partitioning the cluster $c$ is: 
\begin{IEEEeqnarray}{lll}  
\delta M = (\ref{eq:re-mod2}) - (\ref{eq:re-mod1}) \nonumber \\ \nonumber  \\ \nonumber  \hspace{1.85em}
= 2\hat{n}_a\cdot\bold{T}\cdot\hat{n}_a + 2\hat{n}_b\cdot\bold{T}\cdot\hat{n}_b  - 2\hat{n}_c\cdot\bold{T}\cdot\hat{n}_c \\ \nonumber  \\  \hspace{2.87em}
-\> {\delta \vec{N}}\cdot\bold{T}\cdot{\vec{N}} -  {\vec{N}}\cdot\bold{T}\cdot{\delta \vec{N}} - {\delta \vec{N}}\cdot\bold{T}\cdot{\delta \vec{N}}. \label{eq:change-mod0} 
\end{IEEEeqnarray}
Since $G(V, E)$ is an undirected graph, note that the adjacency matrix $\bold{T}$ is a symmetric tensor. This means in equation~(\ref{eq:change-mod0}) ${\delta \vec{N}}\cdot\bold{T}\cdot{\vec{N}} = {\vec{N}}\cdot\bold{T}\cdot{\delta \vec{N}}$.  Therefore, 
\begin{IEEEeqnarray}{lll}
\delta M = 2\hat{n}_a\cdot\bold{T}\cdot\hat{n}_a + 2\hat{n}_b\cdot\bold{T}\cdot\hat{n}_b  - 2\hat{n}_c\cdot\bold{T}\cdot\hat{n}_c \nonumber  \\ \nonumber  \\  \hspace{2.87em}
-\> 2{\vec{N}}\cdot\bold{T}\cdot{\delta \vec{N}} - {\delta \vec{N}}\cdot\bold{T}\cdot{\delta \vec{N}} \nonumber \\ \nonumber  \\ 
\hspace{1.7em}= 2\Delta I_c - 2 {\vec{N}}\cdot\bold{T}\cdot{\delta \vec{N}} - {\delta \vec{N}}\cdot\bold{T}\cdot{\delta \vec{N}},\label{eq:change-mod} 
\end{IEEEeqnarray}
\begin{IEEEeqnarray}{lll}
\hspace{-3.6em}$where~$ \Delta I_c = \hat{n}_a\cdot\bold{T}\cdot\hat{n}_a + \hat{n}_b\cdot\bold{T}\cdot\hat{n}_b  - \hat{n}_c\cdot\bold{T}\cdot\hat{n}_c.~\label{eq:deltaI1}
\end{IEEEeqnarray}
As we mentioned earlier, our objective here is to find a partition that maximizes $\delta M$.  In order to do this, we first derive the results for ${\delta \vec{N}}\cdot\bold{T}\cdot{\delta \vec{N}}$ and $\Delta I_c$, which are on the RHS of $\delta M$~(\ref{eq:change-mod}).  Starting with the expansion of ${{\delta \vec{N}}\cdot\bold{T}\cdot{\delta \vec{N}}}$, given $\delta \vec{N} = \hat{n}_a + \hat{n}_b - \hat{n}_c$~(\ref{eq:deltaN}), we have:
\begin{IEEEeqnarray}{lll}
{{\delta \vec{N}}\cdot\bold{T}\cdot{\delta \vec{N}}} = (\hat{n}_a + \hat{n}_b - \hat{n}_c)\cdot\bold{T}\cdot(\hat{n}_a + \hat{n}_b - \hat{n}_c) \nonumber \\ \nonumber  \\ \nonumber  \hspace{5.3em}
= \hat{n}_a\cdot\bold{T}\cdot\hat{n}_a + \hat{n}_b\cdot\bold{T}\cdot\hat{n}_b  + 2 \hat{n}_a\cdot\bold{T}\cdot\hat{n}_b  \\ \nonumber  \\   \hspace{6.43em}
+\> \hat{n}_c\cdot\bold{T}\cdot\hat{n}_c - 2 \hat{n}_c\cdot\bold{T}\cdot(\hat{n}_a + \hat{n}_b).  ~\label{eq:deltan0}
\end{IEEEeqnarray}
Using $\Delta I_c$ (\ref{eq:deltaI1}) on the RHS of the above equation, we get:
\begin{IEEEeqnarray}{lll}
{\delta \vec{N}}\cdot\bold{T}\cdot{\delta \vec{N}} =  \Delta I_c + 2 \hat{n}_a\cdot\bold{T}\cdot\hat{n}_b - 2\hat{n}_c\cdot\bold{T}\cdot(\hat{n}_a + \hat{n}_b - \hat{n}_c)\nonumber  \\ \nonumber  \\  \hspace{5.3em}
= \Delta I_c + 2 \hat{n}_a\cdot\bold{T}\cdot\hat{n}_b - 2\hat{n}_c\cdot\bold{T}\cdot{\delta \vec{N}}.~\label{eq:deltan1} 
\end{IEEEeqnarray}
On substituting the above result for ${\delta \vec{N}}\cdot\bold{T}\cdot{\delta \vec{N}}$ in equation~(\ref{eq:change-mod}) of $\delta M$, we obtain: 
\begin{IEEEeqnarray}{lll}
\delta M = \Delta I_c - 2 \hat{n}_a\cdot\bold{T}\cdot\hat{n}_b - 2(\vec{N} - \hat{n}_c)\cdot\bold{T}\cdot{\delta \vec{N}}.~\label{eq:deltaMf}
\end{IEEEeqnarray}
To determine $\Delta I_c$ for $\delta M$, given the definition (\ref{eq:deltaI1}), we need to derive results for ${\hat{n}_a\cdot\bold{T}\cdot\hat{n}_a}$, ${\hat{n}_b\cdot\bold{T}\cdot\hat{n}_b}$ and ${\hat{n}_c\cdot\bold{T}\cdot\hat{n}_c}$.  To obtain these results, consider the following generic expansion of ~${\hat{u}\cdot\bold{T}\cdot\hat{v}}$, where the unit vectors $\hat{u}, \hat{v} \in \R^{|V|}$.  From the principles of vector and tensor algebra,
\begin{IEEEeqnarray}{lll}
\hat{u}.\bold{T}.\hat{v} = \sum_{j, k}{T_{jk}}{u_j}{v_k}  \nonumber  \\ \nonumber  \\ \nonumber \hspace{2.8 em}
= \sum_{j, k~\in~a}{T_{jk}}{u_j}{v_k}  + \sum_{j, k~\in~b}{T_{jk}}{u_j}{v_k}   \\ \nonumber  \\ \nonumber \hspace{4.5 em}
+\>  \sum_{j \in a,~k \in b}{T_{jk}}{u_j}{v_k} + \sum_{j \in b,~k \in a}{T_{jk}}{u_j}{v_k} \\ \nonumber  \\  \hspace{4.5 em}
+\> \sum_{j, k~\notin~a\cup b}{T_{jk}}{u_j}{v_k}. ~\label{eq:generic}
\end{IEEEeqnarray}
\\
Given the definition of $\hat{n}_a,~\hat{n}_b$ or $\hat{n}_c$ in equation~(\ref{eq:unitvector}), using the above generic expansion (\ref{eq:generic}), the terms ${\hat{n}_a\cdot\bold{T}\cdot\hat{n}_a}$, ${\hat{n}_b\cdot\bold{T}\cdot\hat{n}_b}$ and ${\hat{n}_c\cdot\bold{T}\cdot\hat{n}_c}$ are determined as:
\begin{IEEEeqnarray}{lll}
\hspace{-1em}{\hat{n}_a\cdot\bold{T}\cdot\hat{n}_a} =  \sum_{j, k~\in~a}{T_{jk}}\frac{1}{\sqrt{n_a}}\frac{1}{\sqrt{n_a}}
=  \frac{1}{n_a}\sum_{j, k~\in~a}{T_{jk}}, ~\label{eq:nana}\\ \nonumber  \\ 
\hspace{-1em} {\hat{n}_b\cdot\bold{T}\cdot\hat{n}_b} =  \frac{1}{n_b}\sum_{j, k~\in~b}{T_{jk}}, \label{eq:nbnb}\\ \nonumber  \\ \nonumber
\hspace{-1em} \&~\hat{n}_c\cdot\bold{T}\cdot\hat{n}_c =\frac{1}{n_c} \sum_{j, k~\in~a}{T_{jk}} + \frac{1}{n_c} \sum_{j, k~\in~b}{T_{jk}} \\ \nonumber  \\ \hspace{6.3 em}
\hspace{-1em}+\>  \frac{1}{n_c} \sum_{j \in a,~k \in b}{T_{jk}}  + \frac{1}{n_c} \sum_{j \in b,~k \in a}{T_{jk}},\label{eq:ncnc}
\end{IEEEeqnarray}
respectively.  Hence, from the definition of ${\Delta I_c}$~(\ref{eq:deltaI1}), we acquire:
\begin{IEEEeqnarray}{lll}
\Delta I_c = (\ref{eq:nana}) + (\ref{eq:nbnb}) - (\ref{eq:ncnc}) \nonumber \\ \nonumber  \\ \nonumber \hspace{1.9 em}
=\Big(\frac{1}{n_a} - \frac{1}{n_c}\Big) \sum_{j, k~\in~a}{T_{jk}}  + \Big(\frac{1}{n_b} - \frac{1}{n_c}\Big) \sum_{j, k~\in~b}{T_{jk}}  \\ \nonumber  \\  \hspace{3.2 em}
-\>  \frac{1}{n_c} \sum_{j \in a,~k \in b}{T_{jk}}  - \frac{1}{n_c} \sum_{j \in b,~k \in a}{T_{jk}} .
\end{IEEEeqnarray}
\\
Since $n_a + n_b = n_c$ (\ref{eq:bisect}), the above expression for $\Delta I_c$ reduces to: 
\begin{IEEEeqnarray}{lll}
\Delta I_c = \frac{n_b}{{n_c}{n_a}} \sum_{j, k~\in~a}{T_{jk}}  + \frac{n_a}{{n_c}{n_b}}\sum_{j, k~\in~b}{T_{jk}} \nonumber \\ \nonumber  \\  \hspace{3.0 em}
-\>  \frac{1}{n_c} \sum_{j \in a,~k \in b}{T_{jk}}  - \frac{1}{n_c} \sum_{j \in b,~k \in a}{T_{jk}}.~\label{eq:deltaI2} 
\end{IEEEeqnarray}

In order to derive relations with the normalized cut approach~\cite{s00}, which we mentioned earlier at the beginning of this section, we need to express $\Delta I_c$  and $\delta M$ in terms of a graph Laplacian.  To do this, we define a unit vector ${\hat{f} = [ f_j] \in {\R}^{n_c}}$, such that $\hat{f}$ is perpendicular to the ones vector ${\bold{1}} \in {\R}^{n_c}$ and the elements of $\hat{f}$ are:
\\
\begin{IEEEeqnarray}{lll}
f_j = \begin{cases}
\sqrt{\frac{n_b}{n_c n_a}} \hspace{1.0 em} $~if~$ j \in a\\ \\
-\sqrt{\frac{n_a}{n_c n_b}} \hspace{0.3 em}  $~if~$ j \in b. \\ 
\end{cases} \label{eq:fj}
\end{IEEEeqnarray}
\\
Let ${\bold{T^c} \in {\R}^{n_c} \times {\R}^{n_c}}$ be a second-order tensor representing the adjacency matrix of the subgraph induced by the cluster $c \in C$, \emph{i.e.}
\begin{IEEEeqnarray}{lll}
\bold{T^c} = [T_{jk}] \Big|_{j, k~\in~c} 
\end{IEEEeqnarray}
\\
From the above definition of $\bold{T^c}$, we deduce the following relations between $\bold{T^c}$ and $\bold{T}$: 
\\
\begin{IEEEeqnarray}{lll}
\sum_{j, k~\in a}{T^c_{jk}} =  \sum_{j, k~\in a}{T_{jk}}~; \sum_{j, k~\in b}{T^c_{jk}} =  \sum_{j, k~\in b}{T_{jk}}~; \nonumber \\ \nonumber \\ \nonumber \hspace{3.0 em} 
 \sum_{j \in a,~k~\in b}{T^c_{jk}} =  \sum_{j \in a,~k~\in b}{T_{jk}}~;   \\ \nonumber \\  \hspace{3.0 em} 
 \sum_{j \in b,~k~\in a}{T^c_{jk}} =  \sum_{j \in b,~k~\in a}{T_{jk}}.~\label{eq:identity}
\end{IEEEeqnarray}
\\
Using the above definitions of $\hat{f}$ and $\bold{T^c}$, consider expanding ${\hat{f}\cdot\bold{T^c}\cdot\hat{f}}$ based on the principles of vector and tensor algebra:
\\
\begin{IEEEeqnarray}{lll}
{\hat{f}\cdot\bold{T^c}\cdot\hat{f}} =  \sum_{j, k~\in~a}{T^{c}_{jk}}{f_j}{f_k}  + \sum_{j, k~\in~b}{T^{c}_{jk}}{f_j}{f_k} \nonumber  \\ \nonumber  \\ \hspace{5.3em}
+\>  \sum_{j \in a,~k \in b}{T^{c}_{jk}}{f_j}{f_k} + \sum_{j \in b,~k \in a}{T^{c}_{jk}}{f_j}{f_k}.~\label{eq:ftf1} 
\end{IEEEeqnarray}
\\
By substituting the values of $f_j$ (\ref{eq:fj}) in the above equation~(\ref{eq:ftf1}), we get: 
\begin{IEEEeqnarray}{lll}
 \nonumber  \\  
{\hat{f}\cdot\bold{T^c}\cdot\hat{f}} = \sum_{j, k~\in~a}{T^{c}_{jk}}\Big({\sqrt{\frac{n_b}{n_c n_a}}}\Big)^2 + \sum_{j, k~\in~b}{T^{c}_{jk}}\Big({\sqrt{\frac{n_a}{n_c n_b}}}\Big)^2 \nonumber  \\  \nonumber  \\ \nonumber \hspace{5.3 em}
+\>  \sum_{j \in a,~k \in b}{T^{c}_{jk}}\Big(-{\sqrt{\frac{n_b}{n_c n_a}}}{\sqrt{\frac{n_a}{n_c n_b}}}\Big)  \\  \nonumber  \\ \nonumber \hspace{5.3 em}
+\>  \sum_{j \in b,~k \in a}{T^{c}_{jk}}\Big(-{\sqrt{\frac{n_a}{n_c n_b}}}{\sqrt{\frac{n_b}{n_c n_a}}}\Big)  \\  \nonumber  \\ \nonumber \hspace{3.0 em}
= \frac{n_b}{{n_c}{n_a}} \sum_{j, k~\in~a}{T^c_{jk}}  + \frac{n_a}{{n_c}{n_b}}\sum_{j, k~\in~b}{T^c_{jk}}  \\ \nonumber  \\  \hspace{5.3 em}
-\>  \frac{1}{n_c} \sum_{j \in a,~k \in b}{T^c_{jk}}  - \frac{1}{n_c} \sum_{j \in b,~k \in a}{T^c_{jk}}.~\label{eq:ftf2} 
\end{IEEEeqnarray}
On implementing the identities presented in (\ref{eq:identity}) and comparing the equations (\ref{eq:ftf2}) and (\ref{eq:deltaI2}), we deduce that: 
\begin{IEEEeqnarray}{lll}
 \Delta I_c = \hat{f}\cdot\bold{T^c}\cdot\hat{f}. 
\end{IEEEeqnarray}
As $\bold{T^c} = \bold{D^c} - \bold{L^c}$, where $\bold{D}^c$ and $\bold{L}^c$ are the degree- and Laplacian matrices~\cite{l07}, respectively, of the subgraph induced by the cluster $c \in C$, we rewrite, 
\begin{IEEEeqnarray}{lll}
 \Delta I_c = \hat{f}\cdot\{\bold{D^c} - \bold{L^c}\} \cdot\hat{f}. \label{eq:laplacian1}
\end{IEEEeqnarray}
Like $\Delta I_c$ in (\ref{eq:laplacian1}), we can also express $\hat{n}_a\cdot\bold{T}\cdot\hat{n}_b$ of $\delta M$ (\ref{eq:deltaMf}) in terms of the graph Laplacian $\bold{L^c}$. To show this, expand $\hat{n}_a\cdot\bold{T}\cdot\hat{n}_b$ using (\ref{eq:generic}) as:
\begin{IEEEeqnarray}{lll}
{\hat{n}_a\cdot\bold{T}\cdot\hat{n}_b} = \frac{1}{\sqrt{n_a n_b}} \sum_{j \in a,~k \in b}{T_{jk}}  + \frac{1}{\sqrt{n_a n_b}} \sum_{j \in b,~k \in a}{T_{jk}}  \nonumber \\ \nonumber  \\  \hspace{3.7em}
= \frac{1}{\sqrt{n_a n_b}} \Big(\sum_{j \in a,~k \in b}{T_{jk}} + \sum_{j \in b,~k \in a}{T_{jk}}\Big). ~\label{eq:nanb}
\end{IEEEeqnarray}
Given $\hat{f}$ (\ref{eq:fj}), note that $\bold{L^c}$ satisfies the following property~\cite{l07}:
\begin{IEEEeqnarray}{lll}
\hat{f}\cdot\bold{L^c}\cdot\hat{f} = \frac{1}{2}\sum_{j, k} T^c_{jk}(f_j - f_k)^2. ~\label{eq:property}
\end{IEEEeqnarray}
On substituting the values of $f_j$ (\ref{eq:fj}) in the above equation, we obtain:
\begin{IEEEeqnarray}{lll}
\hat{f}\cdot\bold{L^c}\cdot\hat{f} = \frac{1}{2} \sum_{j \in a,~k \in b} T^c_{jk} \Big(\sqrt{\frac{n_b}{n_a n_c}} + \sqrt\frac{n_a}{n_b n_c}\Big)^2  \nonumber  \\ \nonumber \\ \nonumber \hspace{5.2em}
+\>  \frac{1}{2} \sum_{j \in b,~k \in a} T^c_{jk} \Big(-\sqrt{\frac{n_a}{n_b n_c}} 
- \sqrt\frac{n_b}{n_a n_c}\Big)^2  \\ \nonumber \\ \nonumber \hspace{4.0em}
= \frac{1}{2}\Big(\frac{n_a}{n_b n_c} + \frac{n_b}{n_a n_c} + \frac{2}{n_c}\Big) \Big(\sum_{j \in a,~k \in b} T^c_{jk} \\ \nonumber \\ \nonumber \hspace{16.2em}
+\>   \sum_{j \in b,~k \in a} T^c_{jk} \Big) \\ \nonumber \\ \nonumber \hspace{4.1em}
=   \frac{(n_a + n_b)^2}{2n_c n_a n_b} \Big(\sum_{j \in a,~k \in b} T^c_{jk} + \sum_{j \in b,~k \in a} T^c_{jk} \Big).
\end{IEEEeqnarray}
As $n_a + n_b = n_c$ (\ref{eq:bisect}),
\begin{IEEEeqnarray}{lll}
\hat{f}\cdot\bold{L^c}\cdot\hat{f} = \frac{n_c}{2{n_a n_b}} \Big(\sum_{j \in a,~k \in b} T^c_{jk} + \sum_{j \in b,~k \in a} T^c_{jk}\Big).~\label{eq:lapalcian2}
\end{IEEEeqnarray}
With the help of the identities in (\ref{eq:identity}), we deduce from (\ref{eq:lapalcian2}) and (\ref{eq:nanb}) that
\begin{IEEEeqnarray}{lll}
\hat{n}_a\cdot\bold{T}\cdot\hat{n}_b = \frac{2\sqrt{n_a n_b}}{n_c}~\hat{f}\cdot\bold{L^c}\cdot\hat{f}. \label{eq:lapalcian3}
\end{IEEEeqnarray}
Finally, on substituting the results acquired for $\Delta I_c$~(\ref{eq:laplacian1}) and $\hat{n}_a\cdot\bold{T}\cdot\hat{n}_b$~(\ref{eq:lapalcian3}) in $\delta M$~(\ref{eq:deltaMf}), we attain:
\begin{IEEEeqnarray}{lll}
\delta M =  \hat{f}\cdot(\bold{D^c} - \bold{L^c})\cdot\hat{f} - \frac{4\sqrt{n_a n_b}}{n_c}~\hat{f}\cdot\bold{L^c}\cdot\hat{f} \nonumber \\ \nonumber \\ \nonumber \hspace{2.9em}
-\> 2(\vec{N} - \hat{n}_c)\cdot\bold{T}\cdot{\delta \vec{N}}  \\ \nonumber \\ \nonumber \hspace{1.8em}
= \overbrace{\hat{f}\cdot\bold{D^c}\cdot\hat{f} \Big[1 - \underbrace{\Big(1 + \frac{4\sqrt{n_a n_b}}{n_c}\Big)\frac{\hat{f}\cdot\bold{L^c}\cdot\hat{f}}{\hat{f}\cdot\bold{D^c}\cdot\hat{f}}}_{\lambda}\Big]}^{{\alpha}} \\ \nonumber \\  \hspace{2.9em}
-\> \underbrace{2(\vec{N} - \hat{n}_c)\cdot\bold{T}\cdot{\delta \vec{N}}}_{{\beta}}. \label{eq:change-mod2}
\end{IEEEeqnarray}
This is a very interesting result, which reveals how maximizing $\delta M$ in the above equation is connected to the normalized cut approach~\cite{s00}.  To interpret the above equation, we begin with the evaluation of $\beta = {2(\vec{N} - \hat{n}_c)\cdot\bold{T}\cdot{\delta \vec{N}}}$ in $\delta M$.  Given the definitions of $\vec{N}$ (\ref{eq:Nsum}) and $\hat{n}_c$ (\ref{eq:unitvector}), all the components of the vector $\vec{N} - \hat{n}_c$ are non-negative as 
\begin{IEEEeqnarray}{lll}
\vec{N} - \hat{n}_c = [N_j - n_{c_j}] = [\sum_{c^{\prime} \in C-c} n_{c^{\prime}_j}] \label{eq:N-nc}
\end{IEEEeqnarray}
and $n_{c^{\prime}_j} \geq 0$ for ${c^{\prime} \in C-c}$.   Likewise, from the definition of $\delta \vec{N}$ (\ref{eq:deltaN}) and given (\ref{eq:bisect}), the components: 
\begin{IEEEeqnarray}{lll}
\delta N_j = \begin{cases}
\frac{1}{\sqrt{n_a}} -  \frac{1}{\sqrt{n_c}} \hspace{1.0 em} $~if~$ j \in a\\ 
\frac{1}{\sqrt{n_b}} -  \frac{1}{\sqrt{n_c}} \hspace{1.0 em} $~if~$ j \in b \\ 
0 \hspace{5.2 em} $~else$.
\end{cases} \label{eq:elements_deltaN}
\end{IEEEeqnarray}
are non-negative.  Also, as $G(V, E)$ has no negative edge weights, all the elements of $\bold{T}$ are non-negative.  Therefore, from (\ref{eq:N-nc}~\&~\ref{eq:elements_deltaN}), the scalar quantity $\beta = {2(\vec{N} - \hat{n}_c)\cdot\bold{T}\cdot{\delta \vec{N}}}$ is always greater than or equal to zero.  Considering that the coefficient of $\beta$ is negative in $\delta M$, we define $\beta$ as the non-negative penalty introduced by the external clusters $c^{\prime} \in C-c$ for bi-partitioning the cluster $c \in C$.  

Regarding $\alpha$ in $\delta M$, $\alpha$ comprises $\bold{D^c},~\bold{L^c}$ and $\hat{f}$.  As we mentioned earlier, $\bold{D}^c$ and $\bold{L}^c$ are  
the degree- and Laplacian matrix representations, respectively, of the subgraph induced by the cluster $c$, and $\hat{f}$ is a unit vector (as in equation \ref{eq:fj}) perpendicular to the ones vector $\bold{1} \in  {\R}^{n_c}$.  Unlike the terms of $\beta$, note that all the terms of $\alpha$ are local to the cluster $c$.  Also in $\alpha$, as $\bold{D}^c$ and  $\bold{L}^c$ are positive- and semi-positive definite matrices, respectively, ${\hat{f} \cdot \bold{D^c} \cdot \hat{f}} > 0$ and ${\hat{f} \cdot \bold{L^c} \cdot \hat{f}} \geq 0$.  This means maximization of $\delta M$ with respect to $\alpha$ (\ref{eq:change-mod2}) requires minimization of $\lambda = \Big(1 + \frac{4\sqrt{n_a n_b}}{n_c}\Big)\frac{\hat{f}\cdot\bold{L^c}\cdot\hat{f}}{\hat{f}\cdot\bold{D^c}\cdot\hat{f}}$, subject to $\hat{f} \perp \bold{1}$ as in~(\ref{eq:fj}).  Such an optimization of $\delta M$ with respect to $\alpha$ is somewhat similar to the normalized cut approach~\cite{s00,l07}, which finds a bi-partition in cluster $c \in C$ by minimizing $\frac{\hat{f}\cdot\bold{L^c}\cdot\hat{f}}{\hat{f}\cdot\bold{D^c}\cdot\hat{f}}$ with $\hat{f} \perp 1$ as in~(\ref{eq:fj}).  However, as our $\delta M$ comprises both $\alpha$ and $\beta$, maximization of $\delta M$ not only requires minimization of the local expression $\lambda = \Big(1 + \frac{4\sqrt{n_a n_b}}{n_c}\Big)\frac{\hat{f}\cdot\bold{L^c}\cdot\hat{f}}{\hat{f}\cdot\bold{D^c}\cdot\hat{f}}$, but also requires minimization of the non-negative penalty $\beta$ introduced by the external clusters $c^{\prime} \in C-c$.  Thus, finding a partition in cluster $c$ by  optimizing $\delta M$ can be seen as a constrained version of the normalized cut approach.  \\

Where the computational complexity of the optimization of $\delta M$ is concerned, like the normalized cut approach~\cite{s00}, finding a partition that best maximizes $\delta M$ is computationally difficult to solve.  Nevertheless, as next steps, our objective is to find an approximation technique to solve the above optimization problem and develop a community detection algorithm based on the maximization of our metric $M$. \\

\section{Conclusions}\label{sec:conclusion}
A new quantitative metric is introduced, by the name of modularity density, for the partitioning of nodes in heterogeneous networks.  Based on the intuitive idea that \emph{a meaningful community is a group of nodes with strong internal associations and weak external associations with nodes of other groups}, our modularity density is mathematically devised for undirected networks with non-negative edge weights.  Maximization of our metric enables community detection with no bias, \emph{i.e.} \emph{no preference for larger clusters over smaller clusters and vice-versa}.  Compared to the versions of modularity density in present literature, our metric allows better detection of weakly separated communities in heterogenous networks.  Moreover, the cost of computing our modularity density is found to be $O(|E|)$, which is on par or better than that of the previous variants. Thus, from all the above characteristics, we conclude that our metric is superior to the previous variants of modularity density in both performance on general cases and computational complexity.  Furthermore, analysis on network partitioning reveals that maximization of our metric has mathematical relations with the minimization of the well-known normalized cut criterion. 

\section*{Acknowledgment}
The authors acknowledge the support provided by the leadership of CKM Analytix for the successful completion of this manuscript.

\end{document}